\documentclass[aps,prb,twocolumn,superscriptaddress,color]{revtex4}
\usepackage{graphicx} 

\usepackage{graphics}
\usepackage{graphicx}
\usepackage{float}
\usepackage{amsfonts}
\usepackage{amssymb}
\usepackage{textcomp}
\usepackage{color}
\usepackage{amsmath}
\usepackage[pdftex,colorlinks=true]{hyperref}
\usepackage{amssymb}
\usepackage{amsthm}
\usepackage{amsfonts}
\usepackage{bbm}
\usepackage{array}

\newcommand{\beqarr}{\begin{eqnarray}}
\newcommand{\eeqarr}{\end{eqnarray}}

\newcommand{\beq}{\begin{equation}}
\newcommand{\eeq}{\end{equation}}

\def\sectionn#1{\noindent\underline{\it #1:}}

\def\tit#1#2#3#4#5{{\em #5}, {#1}{\bf #2}, #3 (#4)}

\def\prl{Phys.\ Rev.\ Lett.\ }
\def\prx{Phys.\ Rev.\ X\ }

\def\prb{Phys.\ Rev.\ B\ }

\def\bei{\begin{itemize}}
\def\eei{\end{itemize}}
\def\beq{\begin{equation}}
\def\eeq{\end{equation}}

\def\cH{{\cal H}}

\begin{document}

\title{Equilibration and Order in Quantum Floquet Matter}

\author{R. Moessner}
\affiliation{Max-Planck-Institut f\"{u}r Physik komplexer Systeme, 01187 Dresden, Germany}

\author{S. L. Sondhi}
\affiliation{Department of Physics, Princeton University, Princeton, New Jersey 08544, USA}

\begin{abstract}
Equilibrium theormodynamics is characterized by two fundamental ideas: 
thermalisation--that  systems approach
a late time thermal state; and phase structure--that thermal states exhibit singular changes as various parameters
characterizing the system are changed. We summarise recent progress 
that has established generalizations of these ideas to periodically
driven, or Floquet, closed quantum systems. This has resulted in the discovery of
entirely new phases which exist only out of equilibrium,
such as the $\pi$-spin glass or Floquet time crystal.
 \end{abstract}

\maketitle

\sectionn{Introduction} Remarkable progress in the physics of closed quantum systems away from equilibrium has occurred over the last decade. This  has been experimental---most strikingly in cold atomic systems,\cite{Bloch08} computational---often involving quantum information  ideas,\cite{schollrev} and intellectual---ranging from a systematic use of entanglement ideas to the long sought demonstration that localization exists in many body systems.\cite{mblreview} Here, we report very recent progress building particularly on the latter, in our understanding  of periodically driven or Floquet many body systems. 

Closed Floquet systems comprise a vast family of  systems 
generally defined by `drives' or time dependent Hamiltonians with $\cH(t+T)=\cH(t)$ for a fixed 
period $T$.
The promise of Floquet systems is that the periodic drive can lead to new physical phenomena, but their peril is the risk of heating 
up to a ``fully scrambled'' or ``infinite temperature'' state, supporting no non-trivial correlations as all configurations occur with the same probability.

The progress reviewed here has established that the peril can  be avoided; that interesting long time steady states can 
be obtained; and that sharply different behaviors can be distinguished and classified, providing 
generalizations of the foundational
thermodynamic notions of 
thermalization and phase structure\footnote{These are generalizations in that they reduce to the familiar ideas in the setting of ergodic,
time dependent Hamiltonian systems.} into the non-equilibrium regime. Indeed, Floquet systems arguably 
represent the maximum known extension 
of equilibrium phase structure in that generic driven systems lacking periodicity are believed to 
heat to infinite temperature. 
Pioneering experiments\cite{bloch-floquetMBL,zhangDTC,ChoiChoi} have very recently started  exploring 
this universe of many body 
Floquet drives.

Our viewpoint is statistical mechanical and restricted to closed/isolated systems. There is also a large and older literature on single particle Floquet systems \cite{Grifoni98} and much recent work on using Floquet physics to engineer non-trivial Hamiltonians as well as on open system physics to use such engineering to interesting ends. We make contact with this larger Floquet universe only where it intersects with our main theme and direct the reader to the literature for this complementary work\cite{potter,rudner,curtgen,curtsp,rahul1,rahul2,chetan1,polkovnikov1,floquetENG,aokioka}.

\sectionn{Floquet basics}
Most broadly, the quantum mechanics of closed systems is concerned with their unitary time evolution governed
by the Schr\"odinger equation ($\hbar=1$)
\beq
\label{eq:unitaryevolution}
i\frac{d}{dt} U(t,t_0)=\cH(t) U(t,t_0)
\eeq
where $U(t,t_0)$ is the unitary time evolution operator that relates states at time $t_0$ to states at time $t$. For 
completely general $\cH(t)$ there is not much else to do than to buckle down and solve (\ref{eq:unitaryevolution}).
For static systems,  $\cH(t) \equiv \cH_0$, life is much simpler 
as $U(t,t_0) = e^{-i (t-t_0)\cH_0 }$, 
and so we learn vast amounts by solving the eigensystem problem for 
$\cH_0$. Specifically, the eigenstates give rise to special, stationary, 
solutions of the Schr\"odinger equation that form a basis
for general time evolution.

The fundamental difference between the Hamiltonians of 
Floquet and static systems is that the latter 
are fully independent of time, 
while the former are only invariant under discrete time translations 
by a period $T$. This difference is analogous to the difference between 
translation invariance of the continuum and of  a lattice. 
There,  the former allows us to study the spectrum of the generator of translations (the momentum)
while the latter requires that we study the spectrum of the discrete translation operator itself, with
states in different bands corresponding to the same quasi-momentum. 
Correspondingly, for Floquet systems one needs to study 
the properties of the family of single period time evolution operators 
$$
U(t_0+T,t_0) = \mathcal{T} e^{-i\int_{t_0}^{t_0+T} dt' \cH(t')}
$$
where $0\leq t_0<T$. 

Let us define $U(T)=U(T,0)$, whose eigenstates
\beq
U(T) |\phi_\alpha\rangle = e^{-i \epsilon_\alpha T} |\phi_\alpha\rangle
\eeq
define special solutions of (\ref{eq:unitaryevolution}), the Floquet eigenstates
\beq|\psi_\alpha (t)\rangle = U(t,0) |\phi_\alpha\rangle \eeq 
which satisfy $|\psi_\alpha (t+T) \rangle = e^{-i \epsilon_\alpha T}|\psi_\alpha (t) \rangle$. Like the stationary solutions of the 
static problem,  they explicitly exhibit the temporal periodicity of the Hamiltonian 
and form a basis for  general time evolution.
The choice of quasienergy $\epsilon_\alpha$ 
is not unique as $\epsilon_\alpha \equiv \epsilon_\alpha +
n_\alpha (2 \pi/T)$. This is the freedom in choosing the operator logrithm in $U(T)=e^{-i H_F T}$, to obtain
what is called the Floquet Hamiltonian $H_F$. 
A final piece of jargon: one refers to a time series spaced $T$ apart as being stroboscopic.

\begin{figure}
\includegraphics[width=0.27\columnwidth]{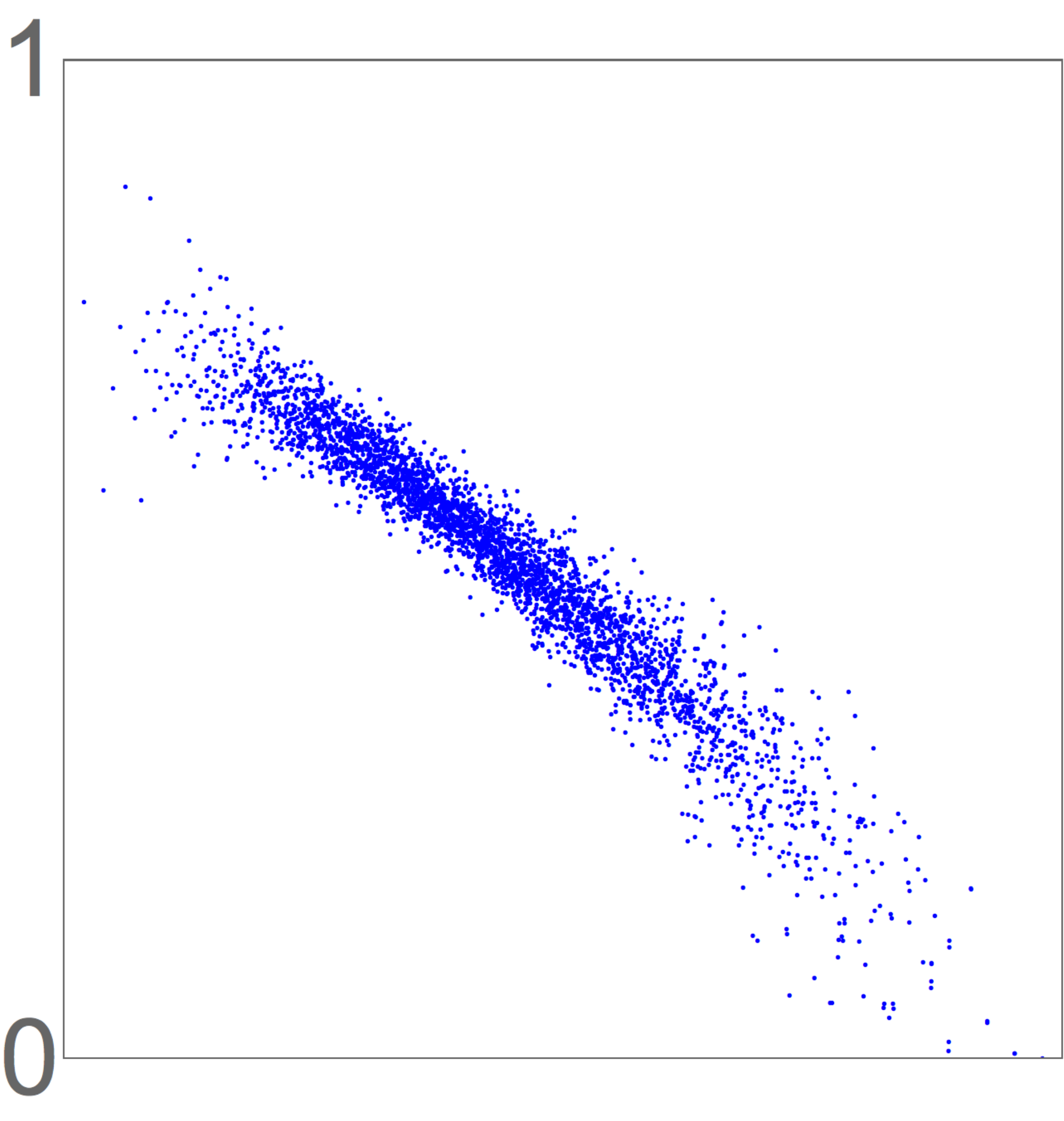}
\includegraphics[width=0.27\columnwidth]{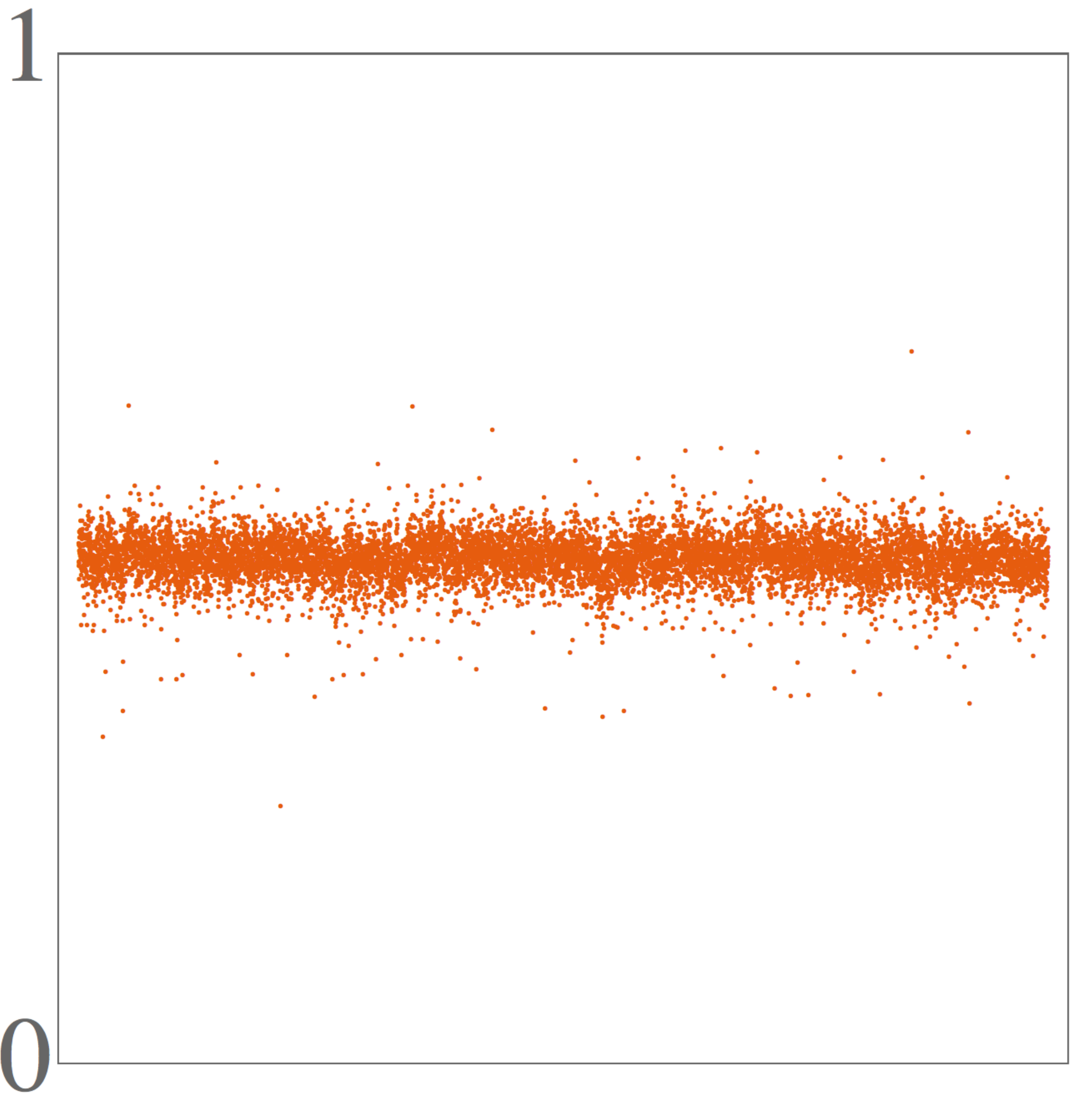}
\includegraphics[width=0.27\columnwidth]{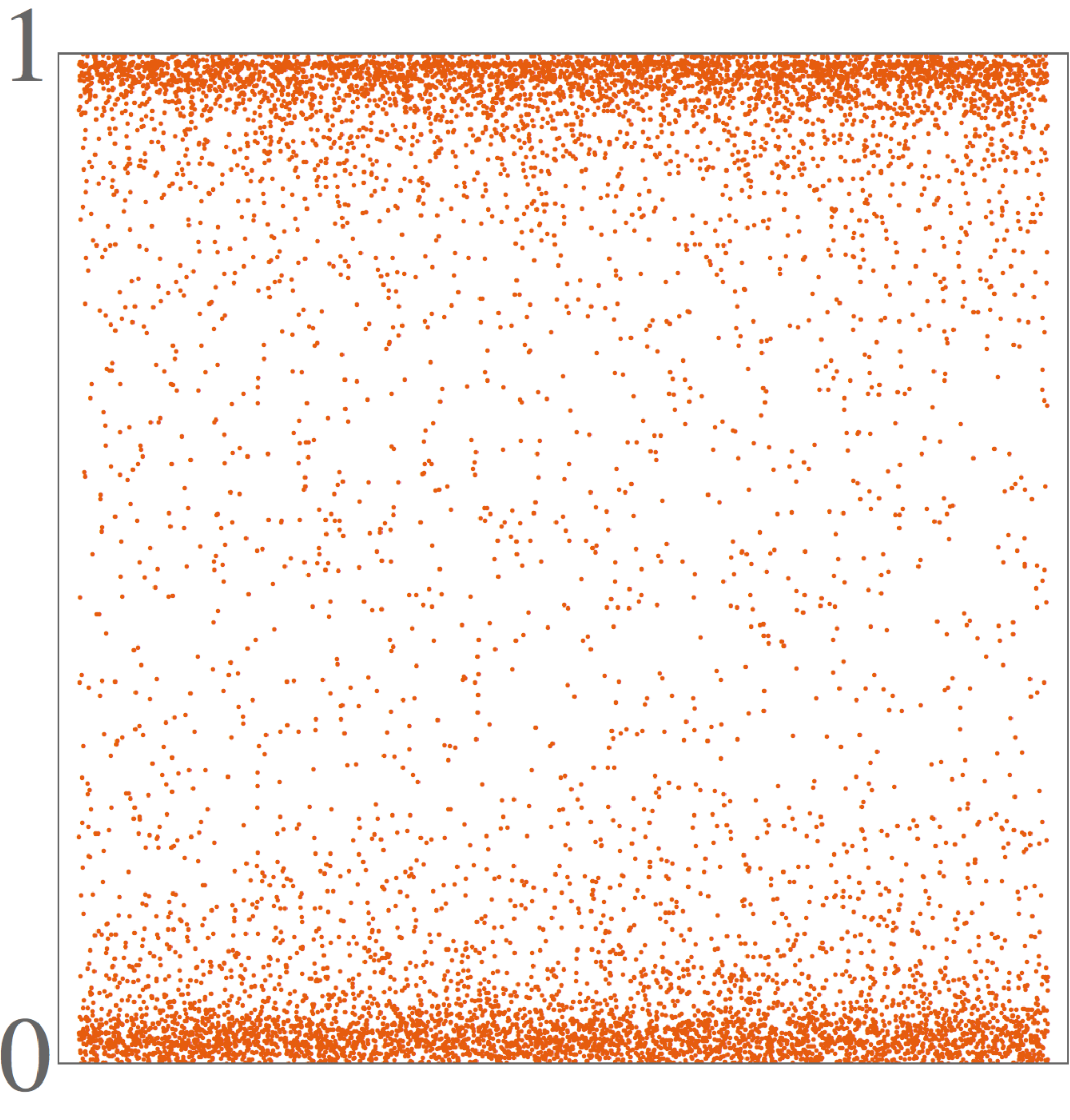}
\caption{Eigenstate properties in a local observable in an inhomogeneous system. (left) Undriven system obeying eigenstate thermalisation: the local observalbe is a monotonic
function of eigenenergy. (middle) Driven system obeying Floquet-ETH: observable is constant function of quasienergy. (right) Driven Floquet-MBL system: observable fluctuates strongly
even between states with adjacent quasienergies.  
} 
\label{eigenstateproperties}
\end{figure}

\sectionn{To heat or not to heat}
We begin with the textbook thermodynamic viewpoint, which notes that systems without continuous time translation symmetry do not conserve 
energy; in particular in periodically driven systems, energy is conserved only modulo $2\pi/T$. 
For generic systems lacking any other local conserved quantities, thermodynamics predicts an entropy maximizing state at late times that is just the 
infinite temperature state \cite{lazarides2,abanin2,refens1}, with all local operator expectation values  time independent at long times irrespective of the
starting state.\footnote{This formulation is not quite crisp (but the conclusion nonethess correct): in a periodically driven system one 
needs to allow for periodic modulation of all quantities and so the proper replacement for time independent steady states
is instead {\it synchronization} in which they all exhibit periodicity with the driving period.} We can reach the same conclusion by noting that linear response theory implies absorption at nonzero frequencies and thus a heating cascade that can only terminate at $T=\infty$.  In this unique ergodic phase, all Floquet eigenstates must individually yield $T=\infty$ correlations and exhibit volume law enanglement with the maximum thermodynamic entropy. This requirement is an incarnation of the eigenstate thermalisation hypothesis (ETH), originally formulated for static ergodic systems \cite{Rigol08} which states that the value of any local observable in an eigenstate is a smooth function of its energy density, as shown in Fig.~\ref{eigenstateproperties}, so that replacing an 
exact eigenstate with an ensemble of states around its energy yields the same thermodynamic behaviour. If the system has a finite number of conserved quantities, other than the now missing energy, the late time states can depend on these. An example would be fermion number for a set of interacting fermions. However given the typically macroscopic number of states in each sector defined by these conserved quantities, we expect that each sector exhibits infinite temperature up to global constraints, although it would be interesting to find examples where the sectors exhibit singular changes as the conserved densities are varied. 

Existence of the Floquet-ergodic phase and applicability of ETH to its Floquet eigenstates has been confirmed computationally. There is considerable evidence that clean, interacting drives generically give rise to this behavior, as  
assumed in the following. However,  exceptions\cite{chansondhi} and apparent exceptions\cite{prosenapp,polkovnikovapp} are known and deserve of more investigation even as there is no good reason to assume that they represent stable behavior.

Leaving such worries aside, the suggestion is  that to avoid 
heating we need $O(N)$ integrals of the of motion, i. e.\ quantities
that commute with $U(T)$, and which can be written as sums of quasi-local terms. 
There are  two known classes of systems where this 
is the case.

The first class is driven free fermion systems\cite{lazarides1} and equivalent interacting spin systems obtained via Jordan-Wigner transformation in $d=1$. 
Such systems are described by a quadratic Floquet Hamiltonian
$$
H_F= \sum_\alpha \epsilon_\alpha a^\dagger_\alpha a_\alpha\ , 
$$
where for $N$ sites, there are $N$ conserved quantities
$$I_\alpha = a^\dagger_\alpha a_\alpha\ .$$ 
For a local $H(t)$, linear combinations of these constants likely always
yield quasi-local conserved quantities. We will return to the implications of this below.

The second class is Floquet systems exhibiting many body localization (MBL). 
Their discovery \cite{abanin1,lazarides3,ponteMBL} came as a byproduct of the explosion of interest in MBL \cite{mblreview}, which
generalizes the venerable Anderson  localization of non-interacting particles to the interacting setting. 
For these systems, it was established that there exists a set of $O(N)$ 
spatially localized, mutually commuting, `l-bit'
 operators $\tau_i^z$ (which depend on details of the drive) such that
$$
[U(T),\tau^z_i]=0 \ .
$$
Floquet-MBL is most intuitive when adding a  weak drive to a static MBL system (although not restricted
to this case). The reference MBL system is itself described by a set of l-bits that commute with its Hamiltonian.
The drive flips groups of l-bits only locally, so that the energy difference between initial and final state is 
bounded above, and it is also 
nonzero as there generically is no local resonance.
Stability of the MBL phase then follows for driving frequencies $2\pi/T$  high compared to the upper bound, 
where 
the system can rather be expected to resemble a set of finite-state Rabi oscillations localized in different 
regions, which does not heat indefinitely. By contrast, for low driving frequencies at fixed driving amplitudes, 
absorbing one (or several) quanta of energy $\omega$  gives rise to transitions between the local levels, 
thereby destroying the MBL phase by local heating. 
A combination of computational 
studies, along with more detailed qualitative and analytic arguments,\cite{ponteMBL,abanin1,lazarides3,huse_rarembl} as well as very recent experimental work,
\cite{bloch-floquetMBL} underpin the belief in the existence of this Floquet-MBL to Floquet-ETH transition.

Note that Floquet-MBL systems avoid heating generically---weak perturbations of Floquet-MBL
drives that leave the period unchanged are also Floquet-MBL. By contrast, free fermion systems
are stable to interactions only when Anderson localized by disorder. 

We next discuss how these systems host generalizations of the two central ideas of thermodynamics -- of equilibrium and phase structure. We take these in reverse order.

\sectionn{Eigenstate Order and Phase structure}
As the Floquet-ETH phase is the only ergodic phase,  all other phases must be localized. To define such phases it is fruitful  to 
generalize the notions of eigenstate order and eigenstate phase transitions from the study of undriven MBL \cite{eigenstateorder} to Floquet systems. Eigenstate order exists when individual many body eigenstates exhibit ordering, of which  the spectrum exhibits
a characteristic signature; at eigenstate phase transitions the eigenstates and eigenvalues can exhibit singular changes as a parameter is varied. \footnote{For static/Floquet erogdic systems, this reduces to the conventional notion of order in the standard ensembles of statistical 
mechanics as nearby/all eigenstates by ETH all yield the same answer.} For Floquet systems order can involve non-trivial variations of the eigenstates 
inside the Floquet period.

To get a sense of how more, and fundamentally new, phases arise,\cite{khemani1} 
we discuss the simplest setting---that of Floquet-MBL chains with an Ising ($\mathbb{Z}_2$) symmetry.
Consider the  binary drive protocol
\begin{equation}
H(t)=
\begin{cases}
 -\sum_s  h_{s}\sigma^x_s  + H_{\rm int} \ \ {\rm \ for} \ 0\leq t<T_1 \\
-\sum_s J_{s}\sigma^z_s \sigma^z_{s+1} +  H_{\rm int} \ \ {\rm \ for} \ T_1\leq t<T
\end{cases}
\end{equation}
where $\sigma^x_s,\sigma^z_s$ are Pauli-matrix operators at site $s$, 
and the $h_{s},J_{s}$ are weakly random about mean values $h$ and $J$ 
 to obtain localization;
the additional interaction terms,  weaker still to preserve localisation, prevent a possible reduction to free fermions. 
All terms commute with a global Ising symmetry $P=\prod_s \sigma^x_s$.

\begin{figure}
\begin{minipage}{3.cm}
\includegraphics[width=\columnwidth]{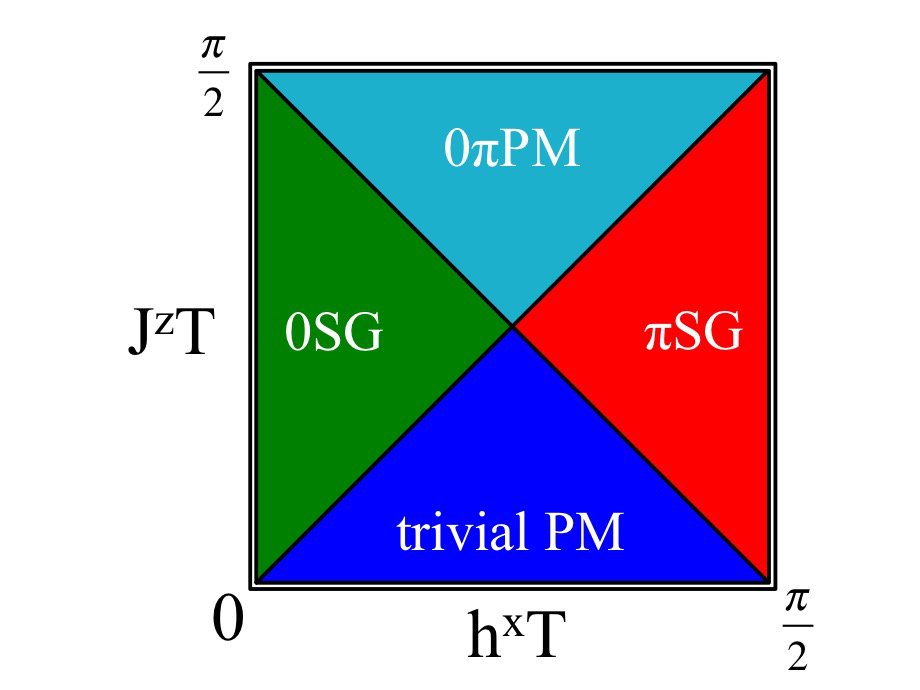}
\end{minipage}
\begin{minipage}{2cm}
\includegraphics[width=0.45\columnwidth]{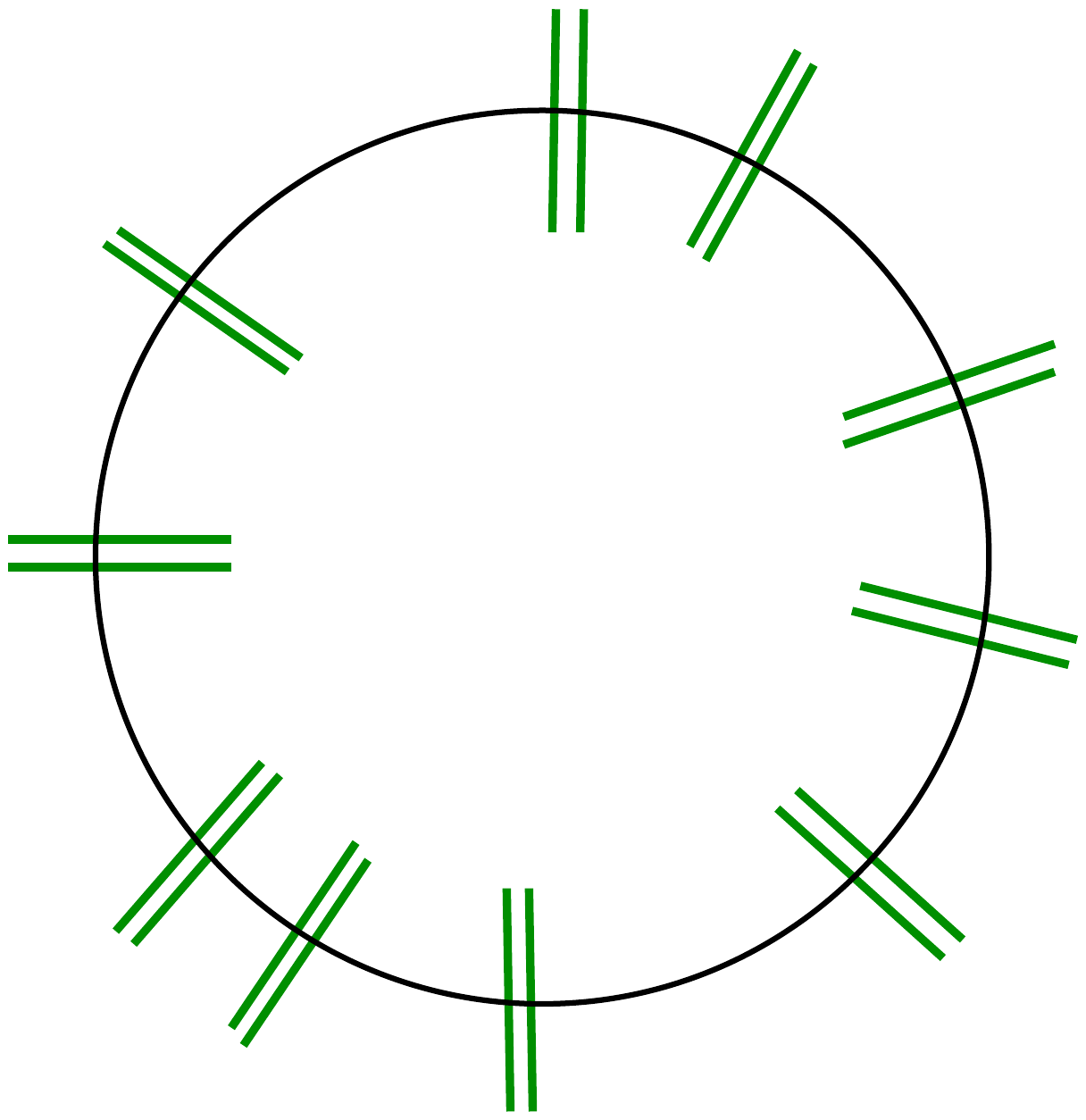}
\includegraphics[width=0.45\columnwidth]{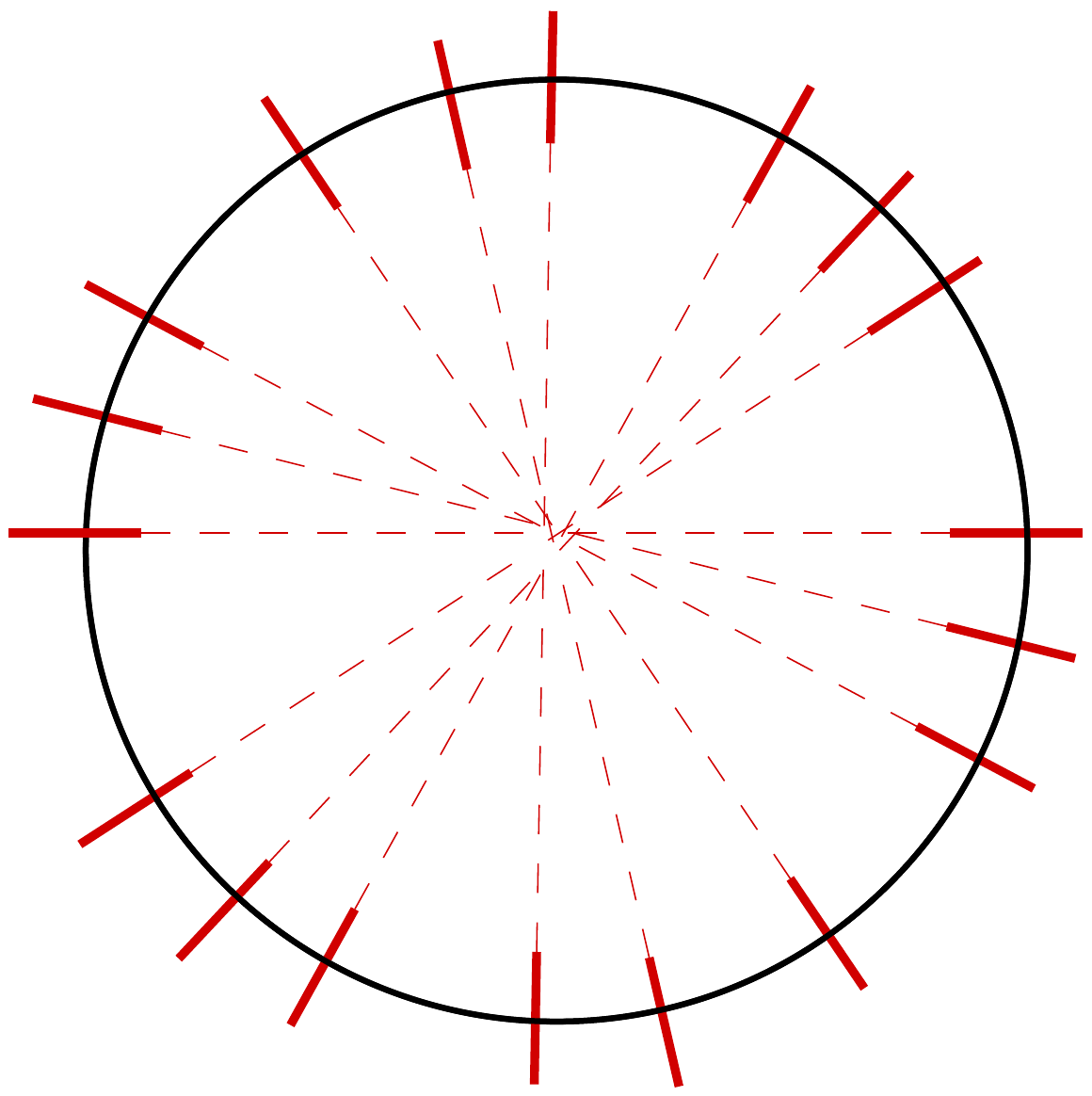}
\includegraphics[width=0.45\columnwidth]{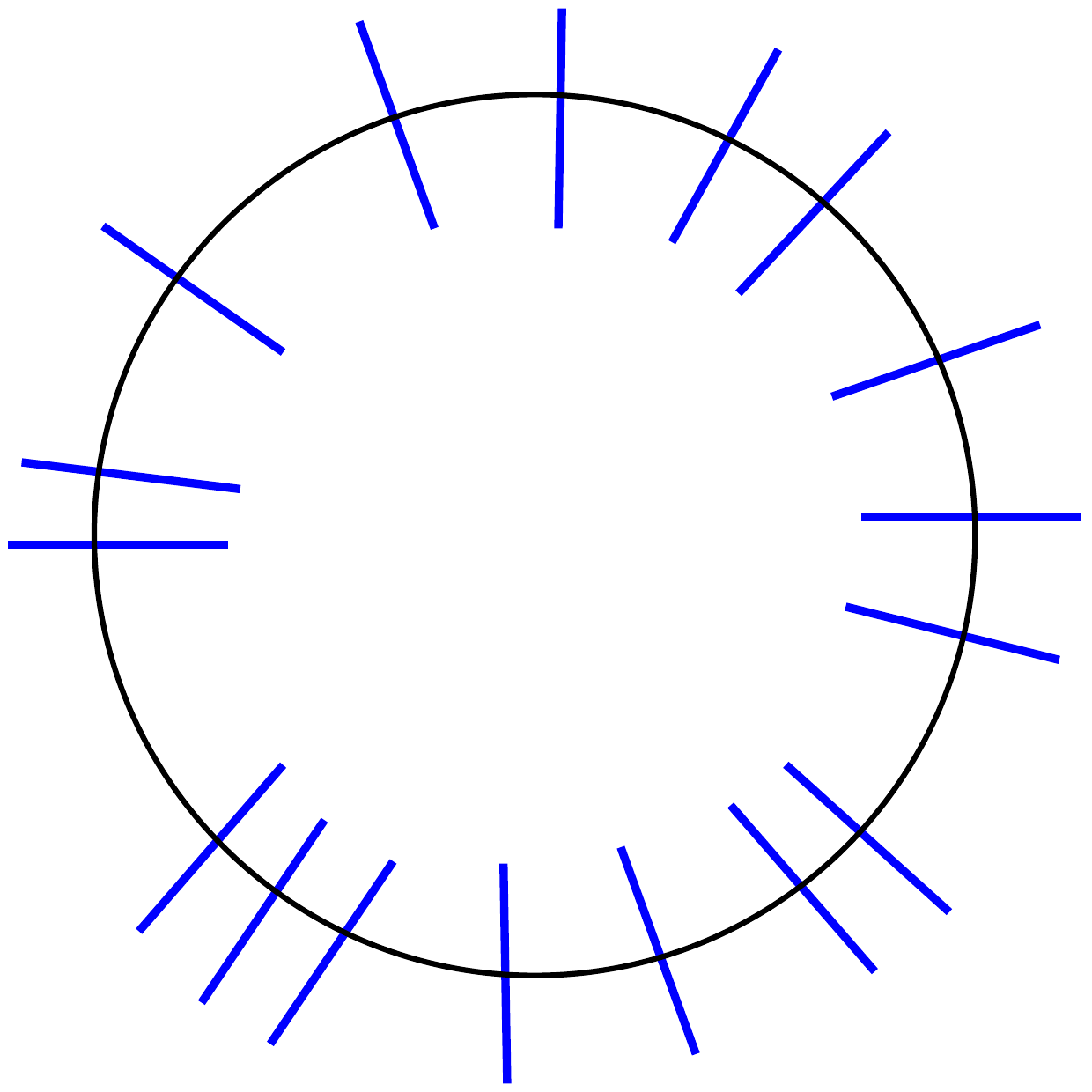}
\end{minipage}
\begin{minipage}{3.cm}
\includegraphics[width=\columnwidth]{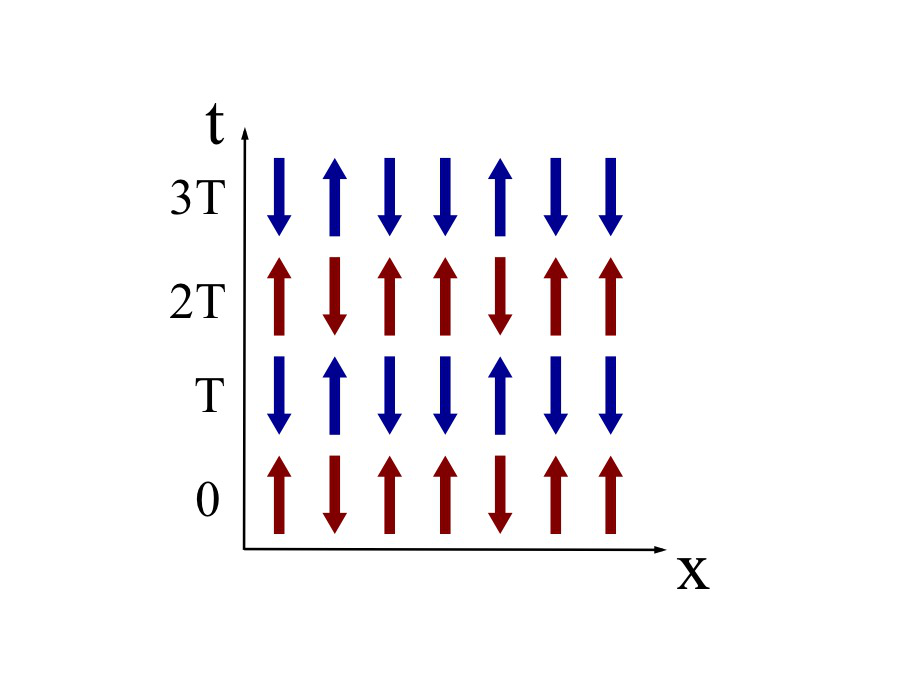}
\end{minipage}
\caption{(left) Phase diagram for the MBL Ising symmetric drives~\cite{khemani1, curtsp} showing $0$SG and $\pi$SG phases which are long-range ordered and spontaneously break Ising symmetry, as well as the $0\pi$-PM and trivial paramagnetic phases without LRO. The $0\pi$PM is an SPT with non-trivial edge modes and can spontaneously break time translation symmetry on its edges. The time crystal $\pi$SG is {\em absolutely stable} in that its existence is not predicated on the Ising symmetry. (middle): Floquet Eigenstate order: the quasienergy axis running from $0$ to $2\pi/T$ is shown as a circle, with location of Floquet eigenenergies shown, which are distributed randomly (PM), in pairs (0SG) and in pairs diametrically separated by $\pi/T$ ($\pi$SG). (right): Period doubling in the $\pi$SG spatiotemporal order as seen in stroboscopic snapshots.
} 
\label{Perturbed}
\end{figure}

This family of drives exhibits exhibits four localized phases. These are shown in the phase diagram Fig~\ref{Perturbed} for the free fermion limit;
with interactions the Floquet ergodic phase will also appear. These phases are characterized as follows in terms of the the spectrum of $U(T)$
and the correlations ${\cal C}_{ij}=\langle \sigma^z_i \sigma^z_j \rangle$ at long
distance $|i-j|\rightarrow\infty$ of the
local Ising odd operators $\sigma^z_i$, Fig.~\ref{Perturbed}:
\begin{itemize}
\item Paramagnet PM (no symmetry-breaking): in all eigenstates ${\cal C}_{ij}\rightarrow 0$.
\item Spin glass SG: in all eigenstates ${\cal C}_{ij} \ne  0$. The spectrum 
contains exponentially degenerate pairs of cat states which are superpositions of states with spin glass order and their
Ising reversed counterparts. Equivalently, in the thermodynamic limit  it consists entirely of states with broken
Ising symmetry and spin glass long range order. Over each period, the order parameter returns to itself as detected by the
depedence of ${\cal C}_{ij}$ {\it within} the period \cite{khemani1}. 
 \item $\pi$-spin glass $\pi$SG: In all eigenstates ${\cal C}_{ij} \ne  0$. The spectrum 
contains pairs of cat states, with splitting is exponentially close to $\pi/T$. 
These are superpositions of states with spin glass order and their Ising reversed counterparts. 
 Even in the thermodynamic limit these cannot be rearranged into states with explicitly
broken Ising symmetry. Thus while the symmetry is broken as indicated by the two-point function, the catness is intrinsic.
Over each period, the order parameter changes sign. 
\item $0\pi$-paramagnet $0\pi$PM: In all eigenstates ${\cal C}_{ij} \rightarrow  0$ {\it in the bulk}. However {\it in open chains}
the spectrum comes in multiplets of four with splittings  exponentially close to 0 and $\pi$; in closed
chains the states are unique. Such phases are known as symmetry-protected topological phases, SPTs: 
trivial in the bulk, but with edge states on open chains. There is
also interesting dynamics at the edge.
\end{itemize}
We emphasize that all these phases exhibit a breakdown of ETH in that the correlators fluctuate strongly between
neighboring eigenstates. Thus, while an average over all states yields $T=\infty$ correlators, individual eigenstates do not, see Fig.~\ref{eigenstateproperties}. 
Also, the eigenstates exhibit area law entanglement which then also serves as an additional eigenstate diagnostic of the
passage between any one of these phases and the ergodic phase. Interestingly, the two new phases can also be classified by
means of local order parameters for time translation symmetry which is generated by $U(T)$ itself. Of these the $\pi$SG breaks 
time-translation symmetry in its bulk, while the $0\pi$PM breaks it only at its boundaries: these provide examples of the ``time crystals''
first hypothesized for undriven systems although the term ``spatio-temporally ordered'' is perhaps more accurate. We
 describe the dynamical consequences of this identification below.

Finally, we note that the $\pi$SG is an exceptionally interesting phase. It is not merely stable to Ising invariant perturbations, instead it is
{\it absolutely stable}\cite{curtabs}---i.e. it is stable to all weak perturbations that do not alter the drive period.
\footnote{A previous paper\cite{chetan2} found a stability axis protected by an anti-unitary Ising symmetry.}
The enlarged phase breaks an emergent Ising symmetry as well as time translation symmetry.

\sectionn{Late time states}
Thus far we have made sharp statements about many body eigenstates. As these are in general not easy to prepare, it is important to 
ask what degree of universality is present in late time states reached by time evolution from more easily prepared initial states; and 
whether the above phases and transitions  can be detected in such late time states. For the ergodic phase, but not for our case, 
ETH ensures that
eigenstate and late time averages agree. Nevertheless,
the late time states are sufficiently robust that the phase structure can indeed be detected. To see this, consider 
 a general state
$$
|\chi(t) \rangle = \sum_\alpha c_\alpha |\psi_\alpha(t) \rangle =  \sum_\alpha c_\alpha e^{-i \epsilon_\alpha t} |\phi_\alpha(t) \rangle
$$
which gives rise to the time dependent expectation value
$$
\langle \chi(t)| O |\chi(t) \rangle =  \sum_\alpha \sum_\beta c_\alpha^* c_\beta e^{-i (\epsilon_\alpha -\epsilon_\beta) t} 
\langle \phi_\beta| O|\phi_\alpha(t) \rangle  \ .
$$
For MBL-Floquet systems $\epsilon_\alpha -\epsilon_\beta$ is essentially continuously distributed in the thermodynamic limit, {\it except}
for the spliitings internal to the spectral multiplets of the kind discussed above. Thus at late times the expectation value reduces to its value in the quasi-diagonal ensemble
\begin{eqnarray}
\lim_{n \rightarrow \infty} \langle \chi(t+nT)| &O& |\chi(t+nT) \rangle \sim \\
 & &\sum_\alpha \sum_{\beta(\alpha)} c_\alpha^\ast c_{\beta(\alpha)} 
 \langle \phi_\alpha(t)| O|\phi_{\beta(\alpha)}(t) \rangle  \ , \nonumber
\end{eqnarray}
so  the late time density matrix is effectively,
$$
\rho \sim \sum_\alpha \sum_{\beta(\alpha)} c_\alpha^\ast c_{\beta(\alpha)}  |\phi_\alpha(t) \rangle \langle  \phi_{\beta(\alpha)}(t)| \ ,
$$
with $\beta(\alpha)$ the member of the multiplet that contains $\alpha$. Thus at late times, roughly half the parameters present in the specification of the initial state (the phases) can no longer be recovered by local measurements.

For the phases of our model Ising drive the following table lists the characteristic behavior of late time states:
\begin{itemize}
\item PM: synchronized and paramagnetic. Expectation values strictly periodic with $T$ with those
of Ising odd operators vanishing for all  starting states.
\item SG: synchronized and break Ising symmetry. For an initial state that breaks Ising symmetry,
one point functions of Ising odd operators are nonzero while for Ising symmetric initial states we need to examine the two
point functions at large distances.
\item $0\pi$PM: synchronized and paramagnetic, except at the boundary, where they exhibit period doubling.
\item $\pi$SG:  break Ising symmetry {\it with period doubling}. For an initial state that breaks Ising 
symmetry, one point functions of Ising odd operators are nonzero while for Ising symmetric initial states we need to examine the two
point functions at large distances. Stroboscopic snapshots look like Fig~\ref{Perturbed}. 
In regions of the $\pi$SG phase lacking a microscopic Ising
symmetry, generic local operators will exhibit period doubling; this has been seen in an experiment.\cite{zhangDTC}
\end{itemize}

Finally we turn to free fermion systems, which turn out to behave differently. Of these, Anderson localized systems share much with
their MBL cousins but they do {\it not} exhibit dephasing and so exhibit late time states with no particular periodicity. 
For free fermion Floquet systems without Anderson localization, stroboscopic evolution with $H_F$ is believed to
lead to late time states which are well captured by a generalized Gibbs ensemble (GGE)
$$
\rho \sim e^{-\sum_\alpha \lambda_\alpha I_\alpha} \ .
$$
With the non-trivial but periodic intra-period evolution included, this has been called the periodic Gibbs ensemble (PGE) or
the Floquet-GGE. It is worth noting that the PGE density matrix leads to a volume law entanglement entropy that is less than 
the infinite temperature value, thus confirming a lack of heating.\cite{sensengupta} 
The moral of this part of the story is that much less information survives in the free fermion late time states than does
in the diagonal ensembles that describe Floquet-MBL systems but more than survives for the Floquet-ETH case.

\sectionn{Recent developments and outlook}
In a flurry of work, the program of identifying stable interacting Floquet phases has been pushed quite far 
already.\cite{curtgen,curtsp,chetan1,rahul1,rahul2} This  builds on an essentially complete classification analogous to that of topological insulators 
and superconductors for free fermion systems.\cite{nathanrudner}
The free fermion classification classifies {\it single-particle} unitaries and
does not always lead to stable many body phases upon the addition of weak interactions as is the case for the analogous question
for undriven free fermion systems.\cite{schnyder} Among the examples which {\it is} stable is the anomalous Floquet
Anderson insulator\cite{AFAI} 
which exhibits chiral edge modes
without delocalized bulk states and is readily realized via a binary drive that appears to be experimentally feasible. 
The free fermion classification is, of course, relevant to experiments that probe few particle physics.

Cold atomic systems, combining long coherence times and tunability of geometry, disorder and interactions, 
provide an ideal platform for testing those ideas. An important development is the demonstration\cite{coldMBL,mbl2d}  of 
(static) MBL in a disordered two-dimensional optical lattice, finding a transition into a regime at which memory of
the initial state with an asymmetric boson occupancy became long-lived. Very recently, an analogous
study\cite{bloch-floquetMBL} was undertaken on a Floquet system with a (quasi-)disorder potential oscillating in time
around a non-zero mean. Here, the memory indicative of  MBL disappears
as the driving frequency is lowered, in keeping with the abovementioned predictions.\cite{abanin1,lazarides3} Finally, a first 
experiment claiming the observation of a discrete time crystal in the time domain has also appeared.\cite{zhangDTC} 
An experimental tour de force, it involves a mesoscopic system, with the experimental verification of the full spatio-temporal order 
in the $\pi$SG remaining an outstanding challenge.

An important line of work that is highly relevant to experiments is on pre-thermal regimes for Floquet systems wherein they
can exhibit plateaux characterized by equilibration with an effective $H_F$ over a long period before finally heating up to
the ergodic steady state.\cite{poti,knap} In principle this makes it possible to observe non-trivial effective phases, such as time crystals,
even in systems that are not localized. Excitingly, a very recent experiment sees such behavior in a three dimensional
system of nitrogen vacancy centers,\cite{ChoiChoi} also in the time domain, although the precise connection to pre-thermalization theory not settled.
There clearly remains much scope for further experimental studies of the increasingly rich and complex phenomena in 
many body Floquet systems.


\begin{thebibliography}{99}

\bibitem{Bloch08}
Immanuel Bloch, Jean Dalibard, Wilhelm Zwerger,
\tit{Rev.\ Mod.\ Phys.\ }{80}{885}{2008}{Many-body physics with ultracold gases}

\bibitem{schollrev}
U. Schollwock, \tit{Ann.\ Phys.}{326}{96}{2011}{The density-matrix renormalization group in the age of matrix product states}

\bibitem{mblreview}
Rahul Nandkishore, David A. Huse
\tit{Annual Review of Condensed Matter Physics}{6}{15-38}{2015}{Many body 
localization and thermalization in quantum statistical mechanics}

\bibitem{bloch-floquetMBL}
Pranjal Bordia, Henrik L\"uschen, Ulrich Schneider, Michael Knap, Immanuel Bloch,
\tit{}{arxiv:1607.07868}{}{}{Periodically Driving a Many-Body Localized Quantum System}

\bibitem{zhangDTC}
J. Zhang, P. W. Hess, A. Kyprianidis, P. Becker, A. Lee, J. Smith, G. Pagano, I.-D. Potirniche, A. C. Potter, A. Vishwanath, N. Y. Yao, C. Monroe,
\tit{}{arXiv:1609.08684}{}{2016}{Observation of a Discrete Time Crystal}

\bibitem{ChoiChoi}
Soonwon Choi, Joonhee Choi, Renate Landig, Georg Kucsko, Hengyun Zhou, Junichi Isoya, Fedor Jelezko, Shinobu Onoda, Hitoshi Sumiya, Vedika Khemani, Curt von Keyserlingk, Norman Y. Yao, Eugene Demler, Mikhail D. Lukin,
\tit{}{arxiv:1610.08057}{}{2016}{Observation of discrete time-crystalline order in a disordered dipolar many-body
system}

\bibitem{Grifoni98}
Milena Grifoni, Peter H\"anggi,
\tit{Phys.\ Rep.\ }{304}{229}{1998}{Driven quantum tunneling}

\bibitem{polkovnikov1}
Marin Bukov, Luca D'Alessio, Anatoli Polkovnikov,
\tit{Advances in Physics\ }{64}{139-226}{2015}{Universal High-Frequency Behavior of Periodically Driven Systems: from Dynamical Stabilization to Floquet Engineering}

\bibitem{potter}
Andrew C. Potter, Takahiro Morimoto, Ashvin Vishwanath
\tit{}{arXiv:1602.05194}{}{}{Topological classification of interacting 1D Floquet phases}

\bibitem{rudner}
Frederik Nathan, Mark S. Rudner,
\tit{}{arXiv:1506.07647}{}{}{Topological singularities and the general classification of Floquet-Bloch systems
Frederik Nathan, Mark S. Rudner }

\bibitem{curtgen}
C. W. von Keyserlingk, S. L. Sondhi,
\tit{}{arXiv:1602.02157}{}{}{Phase Structure of 1d Interacting Floquet Systems I: Abelian SPTs}

\bibitem{curtsp}
C. W. von Keyserlingk, S. L. Sondhi,
\tit{}{arXiv:1602.06949}{}{}{1D Many-body localized Floquet systems II: Symmetry-Broken phases}

\bibitem{rahul1}
Rahul Roy, Fenner Harper,
\tit{}{ arXiv:1602.08089}{}{}{Abelian Floquet SPT Phases in 1D}

\bibitem{rahul2}
Rahul Roy, Fenner Harper
\tit{}{arXiv:1603.06944}{}{}{Periodic Table for Floquet Topological Insulators}

\bibitem{chetan1}
Dominic V. Else, Chetan Nayak
\tit{}{ arXiv:1602.04804}{}{}{Classification of topological phases in periodically driven interacting systems}

\bibitem{floquetENG}
Andre Eckardt, Christoph Weiss, Martin Holthaus, 
\tit{\prl}{95}{260404}{2005}{Superfluid-insulator transition in a periodically driven optical lattice}

\bibitem{aokioka}
Takashi Oka, Hideo Aoki,
\tit{\prb}{79}{081406}{2009}{Photovoltaic Hall effect in graphene}

\bibitem{refens1}
Pedro Ponte, Anushya Chandran, Z. Papic, Dmitry A. Abanin,
\tit{Annals of Physics\ }{353}{196} {2015}{Periodically driven ergodic and many-body localized quantum systems}

\bibitem{abanin2}
Dmitry Abanin, Wojciech De Roeck, Francois Huveneers,
\tit{\prl}{115}{256803}{2015}{Exponentially slow heating in periodically driven many-body systems}

\bibitem{lazarides2}
Achilleas Lazarides, Arnab Das, Roderich Moessner,
\tit{\pre}{90}{012110}{2014}{Equilibrium states of generic quantum systems subject to periodic driving}

\bibitem{Rigol08}
Marcos Rigol, Vanja Dunjko, Maxim Olshanii,
\tit{Nature\ }{452}{854}{2008}{Thermalization and its mechanism for generic isolated quantum systems}

\bibitem{chansondhi}
Anushya Chandran, S. L. Sondhi,
\tit{\prb}{93}{174305}{2016}{Interaction stabilized steady states in the driven O(N) model}

\bibitem{polkovnikovapp}
Roberta Citro, Emanuele G. Dalla Torre, Luca D'Alessio, Anatoli Polkovnikov, Mehrtash Babadi, Takashi Oka, Eugene Demler,
\tit{Annals of Physics\ }{360}{694-710}{2015}{Dynamical Stability of a Many-body Kapitza Pendulum}

\bibitem{prosenapp}
T. Prosen, 
\tit{\prl}{80}{1808}{1998}{Time Evolution of a Quantum Many-Body System: Transition from Integrability to Ergodicity in the Thermodynamic Limit}

\bibitem{lazarides1}
Achilleas Lazarides, Arnab Das, Roderich Moessner,
\tit{\prl}{112}{150401}{2014}{Periodic thermodynamics of isolated systems}

\bibitem{ponteMBL}
Pedro Ponte, Z. Papic, Francois Huveneers, Dmitry A. Abanin,
\tit{\prl}{114}{140401}{2015}{Many-body localization in periodically driven systems}

\bibitem{abanin1}
Dmitry Abanin, Wojciech De Roeck, Francois Huveneers,
\tit{arxiv:}{1412.4752}{}{2014}{A theory of many-body localization in periodically driven systems}

\bibitem{lazarides3}
Achilleas Lazarides, Arnab Das, Roderich Moessner,
\tit{\prl}{115}{030402}{2015}{Fate of many-body localization under periodic driving}


\bibitem{huse_rarembl}
Marin Bukov, Markus Heyl, David A. Huse, Anatoli Polkovnikov,
\tit{\prb}{93}{155132}{2016}{Heating and many-body resonances in a periodically driven two-band system}



\bibitem{eigenstateorder}
David A. Huse, Rahul Nandkishore, Vadim Oganesyan, Arijeet Pal, S. L. Sondhi,
\tit{\prb}{88}{014206}{2013}{Localization protected quantum order}

\bibitem{khemani1}
Vedika Khemani, Achilleas Lazarides, Roderich Moessner, S. L. Sondhi,
\tit{\prl}{116}{250401}{2016}{On the phase structure of driven quantum systems}

\bibitem{curtabs}
C.W. von Keyserlingk, Vedika Khemani, S. L. Sondhi
\tit{}{arXiv:1605.00639}{}{2016}{Absolute Stability and Spatiotemporal Long-Range Order in Floquet systems}

\bibitem{chetan2}
Dominic V. Else, Bela Bauer, Chetan Nayak,
\tit{}{arXiv:1603.08001}{}{2016}{Floquet Time Crystals}


\bibitem{sensengupta}
Arnab Sen, Sourav Nandy, K. Sengupta, \tit{}{arXiv:1511.03668 }{}{}{Entanglement generation in periodically driven integrable systems: dynamical phase transitions and steady state}

\bibitem{nathanrudner}
Frederik Nathan, Mark S. Rudner,
\tit{}{arXiv:1506.07647}{}{}{Topological singularities and the general classification of Floquet-Bloch systems}

\bibitem{schnyder}
Xue-Yang Song, Andreas P. Schnyder, 
\tit{}{arXiv:1609.07469}{}{}{Interaction effects on the classification of crystalline topological insulators and superconductors}

\bibitem{AFAI}
Paraj Titum, Erez Berg, Mark S. Rudner, Gil Refael, Netanel H. Lindner,
\tit{\prx}{6}{021013}{2016}{The anomalous Floquet-Anderson insulator as a non-adiabatic quantized charge pump}

\bibitem{coldMBL}
Pranjal Bordia, Henrik P. L\"{u}schen, Sean S. Hodgman, Michael Schreiber, Immanuel Bloch, Ulrich Schneider,
\tit{\prl}{116}{140401}{2016}{Coupling Identical 1D Many-Body Localized Systems}

\bibitem{mbl2d}
Jae-yoon Choi, Sebastian Hild, Johannes Zeiher, Peter Schau\ss, Antonio Rubio-Abadal, Tarik Yefsah, Vedika Khemani, David A. Huse, Immanuel Bloch, Christian Gross,
\tit{Science\ }{352}{1547}{2016}{Exploring the many-body localization transition in two dimensions}

\bibitem{poti}
Ionut-Dragos Potirniche, Andrew C. Potter, Monika Schleier-Smith, Ashvin Vishwanath, Norman Y. Yao,
\tit{}{arXiv:1610.07611}{}{}{Floquet symmetry-protected topological phases in cold atomic systems}

\bibitem{knap}
Simon A. Weidinger, Michael Knap,
\tit{}{arXiv:1609.09089}{}{}{Floquet prethermalization and regimes of heating in a periodically driven, interacting quantum system}








\end{thebibliography}
 \end{document}